\documentclass[pra,twocolumn,showpacs,floatfix,a4paper,superscriptaddress]{revtex4}
\usepackage{graphicx}
\usepackage{amsmath}
\usepackage{amssymb}
\usepackage{amsfonts}
\usepackage{color}

\begin{document}

\title{Quantum dynamics of an optical cavity coupled to a thin semi-transparent membrane: effect of membrane absorption}

\author{C. Biancofiore}
 \affiliation{School of Science and Technology, Physics Division, University of Camerino, via Madonna delle Carceri, 9, I-62032 Camerino (MC), Italy, and INFN, Sezione di Perugia, Italy}
\author{M. Karuza}
\affiliation{School of Science and Technology, Physics Division, University of Camerino, via Madonna delle Carceri, 9, I-62032 Camerino (MC), Italy, and INFN, Sezione di Perugia, Italy}
\author{M. Galassi}
 \affiliation{School of Science and Technology, Physics Division, University of Camerino, via Madonna delle Carceri, 9, I-62032 Camerino (MC), Italy, and INFN, Sezione di Perugia, Italy}
\author{R. Natali}
\affiliation{School of Science and Technology, Physics Division, University of Camerino, via Madonna delle Carceri, 9, I-62032 Camerino (MC), Italy, and INFN, Sezione di Perugia, Italy}
\author{P. Tombesi}
 \affiliation{School of Science and Technology, Physics Division, University of Camerino, via Madonna delle Carceri, 9, I-62032 Camerino (MC), Italy, and INFN, Sezione di Perugia, Italy}
 \affiliation{CriptoCam S.r.l., via Madonna delle Carceri 9, I-62032 Camerino (MC), Italy}
\author{G. Di Giuseppe}
\affiliation{School of Science and Technology, Physics Division, University of Camerino, via Madonna delle Carceri, 9, I-62032 Camerino (MC), Italy, and INFN, Sezione di Perugia, Italy}
\affiliation{CriptoCam S.r.l., via Madonna delle Carceri 9, I-62032 Camerino (MC), Italy}
\author{D. Vitali}
 \affiliation{School of Science and Technology, Physics Division, University of Camerino, via Madonna delle Carceri, 9, I-62032 Camerino (MC), Italy, and INFN, Sezione di Perugia, Italy}

\begin{abstract}
We study the quantum dynamics of the cavity optomechanical system formed by a Fabry-Perot cavity with a thin vibrating membrane at its center. We determine in particular to what extent optical absorption by the membrane hinders reaching a quantum regime for the cavity-membrane system. We show that even though membrane absorption may significantly lower the cavity finesse and also heat the membrane, one can still simultaneously achieve ground state cooling of a vibrational mode of the membrane and stationary optomechanical entanglement with state-of-the-art apparatuses.
\end{abstract}


\maketitle


\section{Introduction}

Cavity optomechanics has recently emerged as a rapidly developing research field in which relevant tasks such as quantum limited
displacement sensing, highly sensitive force measurements, and the realization of quantum interfaces for quantum information networks, may be realized by virtue of the radiation pressure coupling between the light confined in a cavity and nano- and micro-mechanical resonators \cite{kippenberg,amo,marquardt0}. The standard and simplest optomechanical setup is a Fabry-Perot cavity in which one of the two mirrors is a vibrating micromechanical object. It has been the first setup experimentally studied \cite{cohadon99,gigan06,arcizet06b,bouwm,mavalvala}, and for which ground state cooling of the resonator has been recently approached \cite{markus09}, and strong optomechanical coupling has been demonstrated \cite{groblacher}. In order to enter in a regime where the dynamics is distinctly quantum, the optomechanical coupling has to dominate both optical and mechanical losses, which implies a high-finesse optical cavity enabling to amplify the radiation pressure, and a movable mirror with high mechanical quality factor $Q_m$. However, it is technically non-trivial to fabricate devices possessing simultaneously high $Q_m$ and high reflectivity, and therefore novel devices have been recently designed and tested. A first example is provided by silica toroidal optical microresonators, coupled by the radiation pressure of the circulating light with the radial vibrational modes of the support \cite{kippenberg,armani,vahalacool,sidebcooling}. Other novel approaches propose to avoid completely clamping mechanical losses by optically trapping levitating dielectrics as in \cite{pnas}. Furthermore, other optomechanical devices have been tested, in which the coupling is provided by the transverse gradient force, such as suspended silicon photonic waveguides \cite{hong}, SiN nanowire evanescently coupled to a microtoroidal resonator \cite{anets}, and a "zipper" cavity formed by two adjacent photonic crystal wires \cite{eichenfeld}.

A further solution has been proposed in Refs.~\cite{harris,njp} in which the mechanical degree of freedom is represented by a partially reflective thin vibrating membrane inserted within a standard cavity with two bulk mirrors. A high value of $Q_m$ is reached by using commercially available high-stress SiN membranes, which are known to possess very low intrinsic losses \cite{zwickl}. At the same time one has a high-finesse cavity by using standard highly reflective mirrors and minimizing optical absorption in the membrane, which can be achieved using stoichiometric SiN \cite{kimble2} and very thin membranes. As recently shown in Ref.~\cite{harris2}, this ``membrane-in-the-middle'' setup has the further advantage that the radiation pressure coupling can be tuned \emph{in situ} by adjusting the position and orientation of the membrane within the cavity \cite{harris2}. In particular optomechanical couplings which are nonlinear in the resonator position can be engineered and tuned. In this paper we study this cavity-membrane system by investigating to which extent optical absorption by the membrane affects reaching the quantum regime. We shall see that even if absorption appreciably decreases the cavity finesse and also heats the membrane, it is still possible to find realistic parameter settings where quantum behavior of the cavity-membrane system is detectable, even up to room temperature.

In Sec.~II we describe the system by means of quantum Langevin equations, showing in particular how the dynamics is modified by optical absorption by the membrane. In Sec.~III we focus on the quantum behavior of the optomechanical system, and show that ground state cooling of the mechanical resonator, and steady-state optomechanical entanglement are achievable even for a non-negligible absorption. Sec.~IV is for concluding remarks, while details of the derivation of the standard single-mode cavity optomechanics description from the general multimode radiation pressure interaction are given in the Appendix.

\section{Description of the system}

\begin{figure}[h]
   \centering
   \includegraphics[width=.45\textwidth]{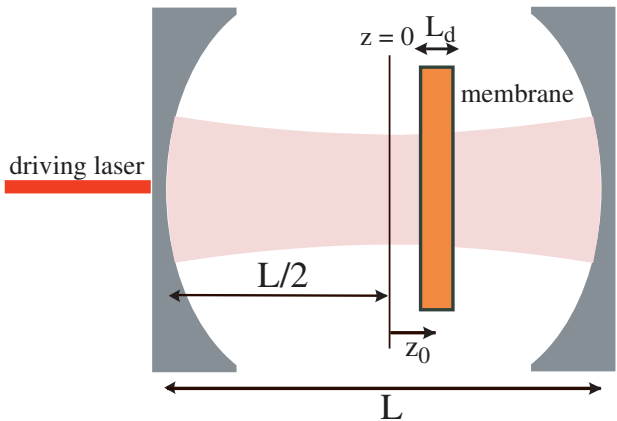}
   \caption{(Color online) Schematic description of the cavity formed by two identical mirrors, with a thin semi-transparent vibrating membrane placed inside.}
   \label{fig:scheme}
\end{figure}

We consider a thin semi-transparent dielectric membrane placed within a driven high-finesse Fabry-Perot cavity of length $L$ (see Fig.~\ref{fig:scheme}). The membrane vibrational modes are coupled to the optical cavity modes by radiation pressure, and therefore one has in general a multimode bosonic system in which mechanical and optical modes interact in a nonlinear way. However, in most practical situations, one can adopt a simplified description based on a \emph{single} cavity mode interacting with a \emph{single} mechanical mode~\cite{harris,njp}. One can restrict to a single cavity mode whenever the driving laser populates a given cavity mode only (typically a TEM$_{00}$ mode, here associated with the annihilation operator $\hat{a}$), and if scattering from the driven mode to other modes is negligible. This latter condition is satisfied when the frequencies of the membrane vibrational modes $\Omega_i$ are smaller than the typical cavity mode separation, which is of the order of the free spectral range $c/2L$ (see also Ref.~\cite{law}). Moreover, one can consider a single mechanical mode of the membrane (described by dimensionless position $\hat{q}$ and momentum $\hat{p}$ operators, such that $[\hat{q},\hat{p}]= \imath$) when the detection bandwidth is chosen so that it includes only a single, isolated, mechanical resonance with frequency $\Omega_m$. In Appendix \ref{bigappen} the general multimode description of the membrane-in-the-middle setup is provided and we discuss the conditions under which such a single-mode reduction is possible.

By explicitly including cavity driving by a laser with frequency $\omega_L$ and input power ${\mathcal P}$, one ends up with the following cavity optomechanical Hamiltonian,
\begin{equation}
H=\frac{\hbar \Omega _{m}}{2}(\hat{p}^{2}+\hat{q}^{2})+\hbar \omega(\hat{q})\hat{a}^{\dagger }\hat{a}
+i\hbar E(a^{\dagger }e^{i\omega_L t}-ae^{-i\omega_L t}),  \label{eq:Ham-optomech}
\end{equation}
where $E=\sqrt{2\mathcal{P}\kappa_0 /\hbar \omega _{L}}$, with $\kappa_0$ the cavity mode bandwidth in the absence of the membrane.
In Eq.~(\ref{eq:Ham-optomech}) we have included the radiation pressure interaction within the cavity energy term by defining the position-dependent frequency
\begin{equation}\label{eq:pos-dep-freq-gen}
\omega(\hat{q})=\omega_0+\delta\omega\left[z_0(\hat{q})\right],
\end{equation}
where $\omega_0$ is the cavity mode frequency in the absence of the membrane, and $\delta\omega\left[z_0(\hat{q})\right]$ is the frequency shift caused by the insertion of the membrane (see Eq.~(\ref{eq:freq-membr-shift}) for its explicit form and Appendix A for its derivation). This shift depends upon the membrane position along the cavity axis $z_0(\hat{q})$, which in turn depends upon the coordinate $\hat{q}$ because one can write $z_0(\hat{q}) = z_0+x_0 \hat{q}$, where $z_0$ is the membrane center-of-mass position along the cavity axis (see Fig.~1) and $x_0=\sqrt{\hbar/m \Omega_m}$, with $m$ the effective mass of the mechanical mode \cite{harris,njp,kimble2}.

A first order expansion in $\hat{q}$ of Eq.~(\ref{eq:pos-dep-freq-gen}) provides an accurate description of the physics in most cases, except when the membrane center $z_0$ is placed exactly at a node or at an antinode of the cavity field, where the first-order term in the expansion of $\omega(\hat{q})$ vanishes, and one has to consider the higher-order term, which is quadratic in $\hat{q}$. This latter term describes a dispersive interaction between the optical and the vibrational modes which has been discussed in \cite{harris,harris2,bhatta,nunnekamp} and may be exploited for a quantum non-demolition measurement of the vibrational energy.

\subsection{Quantum Langevin description of dissipation and noise sources}

The dynamics are determined not only by the Hamiltonian of Eq.~(\ref{eq:Ham-optomech}), but also by the fluctuation-dissipation processes affecting both the optical and the mechanical mode. The mechanical mode is affected by a viscous force with damping rate $\gamma_m$ and by a Brownian stochastic
force with zero mean value $\hat{\xi}(t)$, obeying the correlation function \cite{Landau,gard,GIOV01}
\begin{equation}  \label{eq:browncorre}
\left \langle \hat{\xi}(t) \hat{\xi}(t^{\prime})\right \rangle = \frac{\gamma_m}{\Omega_m%
} \int \frac{d\omega}{2\pi} e^{-i\omega(t-t^{\prime})} \omega \left[%
\coth\left(\frac{\hbar \omega}{2k_BT_0}\right)+1\right],
\end{equation}
where $k_B$ is the Boltzmann constant and $T_0$ is the temperature of the reservoir of the membrane. The Brownian noise $\hat{\xi}(t)$
is a non-Markovian, Gaussian quantum stochastic process, but in the limit of high mechanical
quality factor $Q_m=\Omega_m/\gamma_m \gg 1$, becomes with a good approximation Markovian, with correlation function \cite{gard,GIOV01}
\begin{equation}
\label{eq:browncorre2}\left\langle \hat{\xi}(t)\hat{\xi}(t^{\prime})\right\rangle \simeq \gamma_{m}\left[(2n_{0}+1) \delta(t-t^{\prime})+i
\frac{\delta^{\prime}(t-t^{\prime})}{\Omega_{m}}\right]  ,
\end{equation}
where $n_{0}=\left( \exp \{\hbar \Omega _{m}/k_{B}T_{0}\}-1\right) ^{-1}$ is the mean thermal phonon number at $T_{0}$, and $\delta^{\prime}(t-t^{\prime})$ denotes the derivative of the Dirac delta.

Then, due to the nonzero transmission of the cavity mirrors, the cavity field decays at rate $\kappa_0 $ and is affected by the vacuum optical input noise $\hat{a}_0^{in}(t)$, whose only nonzero correlation functions is given by \cite{gard}
\begin{equation}\label{eq:voinoise}
 \langle \hat{a}_0^{in}(t)\hat{a}_0^{in,\dag }(t^{\prime })\rangle = \delta (t-t^{\prime }).
\end{equation}
However, the presence of the membrane and of its optical absorption provides an additional loss channel for the cavity photons, which has been neglected up to now \cite{harris,bhatta}. Absorption by the membrane is quantified by the imaginary part to the the refraction index, $n_M=n_R+\imath n_I$ ($n_I \ll n_R$), yielding a small imaginary part of the frequency shift $\delta\omega\left[z_0(\hat{q})\right]$. As a consequence, the membrane absorbs cavity photons with a rate
\begin{equation}\label{eq:nonl-dec-rate}
\kappa_1(\hat{q})\equiv \left|{\rm Im}\left\{\delta\omega\left[z_0(\hat{q})\right]\right\}\right|,
\end{equation}
which depends upon the effective resonator position operator $\hat{q}$ and therefore describes a nonlinear dissipative process affecting also the mechanical mode. Due to fluctuation-dissipation theorem, this nonlinear dissipation is associated with an additional optical input noise $\hat{a}_1^{in}(t)$, possessing the same autocorrelation functions of $\hat{a}_0^{in}(t)$ [see Eq.~(\ref{eq:voinoise})], and describing the vacuum optical field fluctuations re-radiated within the cavity by the membrane. At the same time this latter noise is responsible for an additional stochastic force on the mechanical mode, describing membrane heating due to absorption.

\begin{figure}[h]
   \centering
   \includegraphics[width=.48\textwidth]{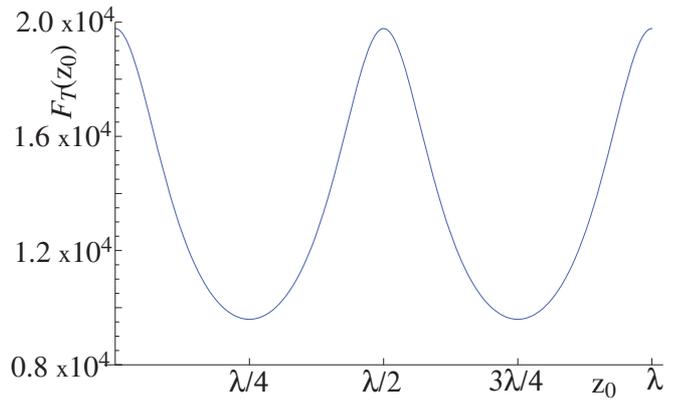}
   \caption{Cavity finesse versus the membrane position along the cavity axis $z_0$. The plot refers to an empty cavity finesse ${\mathcal F}=20000$, and to a membrane with thickness $L_d = 50$ nm, and index of refraction $n_M =2.0+i 10^{-4}$. }
   \label{finesse}
\end{figure}

Therefore, the cavity mode has a total decay rate $\kappa_T(\hat{q})=\kappa_0+\kappa_1(\hat{q})$; this means that the cavity finesse is decreased from its empty cavity value ${\mathcal F}$ by membrane absorption, and that, due to Eq.~(\ref{eq:nonl-dec-rate}), it becomes dependent upon the membrane position $z_0$ according to
\begin{equation}\label{eq:finesse}
    \frac{1}{{\mathcal F}_T(z_0)}=\frac{1}{{\mathcal F}}+\frac{2}{\pi}\left|{\rm Im}\left\{\frac{L \delta \omega(z_0)}{c}\right\}\right|.
\end{equation}
Using the explicit expression of the frequency shift derived in Appendix A, Eqs.~(\ref{eq:freq-membr-shift}), Eq.~(\ref{eq:finesse}) implies that the resulting finesse ${\mathcal F}_T(z_0)$ is a periodic function of $z_0$ with period equal to half wavelength of the mode, with maxima attained when the membrane is centered around a node of the cavity field \cite{harris,njp,kimble2}. Such a behavior, in the case of an empty cavity finesse ${\mathcal F}=20000$ and for $n_I=10^{-4}$, is shown in Fig.~\ref{finesse}, where ${\mathcal F}_T(z_0)$ decreases down to almost $50\%$ of its initial value by varying the membrane position $z_0$.

Adding the above damping and noise terms to the Heisenberg equations of motion associated with the Hamiltonian of Eq.~(\ref{eq:Ham-optomech}), one gets the following set of nonlinear quantum Langevin equations (QLE) which, in
the frame rotating at the driving laser frequency $\omega_{L}$, read
\begin{subequations}
\label{nonlinlang}
\begin{eqnarray}
\dot{\hat{q}}& =& \Omega _{m}\hat{p}, \label{nonlinlang1}\\
\dot{\hat{p}}& = & -\Omega _{m}\hat{q}-\gamma _{m}\hat{p}-\partial_q \omega(\hat{q}) \hat{a}^{\dagger }\hat{a}\nonumber \\
&& +\hat{\xi} -i\frac{\partial_q \kappa_1(\hat{q})}{\sqrt {2\kappa_1(\hat{q})}}\left[\hat{a}^{\dagger}\hat{a}_1^{in}-\hat{a} \hat{a}_1^{in,\dagger}\right], \label{nonlinlang2}\\
\dot{\hat{a}}& = & -i\left[\omega(\hat{q})-\omega_L\right]\hat{a}-\left[\kappa_0+\kappa_1(\hat{q})\right]\hat{a}
+E \nonumber \\
&&+\sqrt{2\kappa_0}\hat{a}_0^{in}+\sqrt{2\kappa_1(\hat{q})}\hat{a}_1^{in}, \label{nonlinlang3}
\end{eqnarray}
\end{subequations}
where $\partial_q$ denotes the derivative with respect to $\hat{q}$. These nonlinear QLE exactly describe the dynamics of the cavity-membrane system, but they are difficult to solve due to the various nonlinearities. Membrane absorption in particular is responsible for the additional nonlinear damping term acting on the cavity mode, and for the additional nonlinear noise term associated with membrane heating.

\subsection{Linearized quantum Langevin equations}

In most practical situations it is not necessary to solve the nonlinear QLE of Eqs.~(\ref{nonlinlang}). In fact, one is interested in parameter regimes where the radiation pressure interaction has appreciable effects and a simple way to achieve it is to have an intense intracavity field, which is achievable with a large cavity finesse ${\mathcal F}$ and with enough driving power. In this case the operation point is given by a classical steady state characterized by a coherent intracavity field with amplitude $\alpha_s$ ($|\alpha_s| \gg 1$) and a deformed membrane mode with a new stationary position $q_s$, satisfying the coupled nonlinear conditions
\begin{subequations}
\label{eq:stat}
\begin{eqnarray}
\label{eq:stat1}
  q_s &=& -\frac{\partial_q \omega(q_s)|\alpha_s|^2}{\Omega_m}, \\
  |\alpha_s|^2 &=& \frac{|E|^2}{ \left[\kappa_0+\kappa_1(q_s)\right]^2+\left[\omega_L-\omega(q_s)\right]^2}. \label{eq:stat2}
\end{eqnarray}
\end{subequations}
Eq.~(\ref{eq:stat2}) in particular is the nonlinear equation determining $|\alpha _{s}|$, due to the dependence of $q_s$ upon $|\alpha _{s}|$ itself given by Eq.~(\ref{eq:stat1}), and which may show optical bistability, i.e., the presence of two simultaneous stable solutions above a given threshold for the input power ${\mathcal P}$, as demonstrated in \cite{dorsel} and also with Bose-Einstein condensates in \cite{brennecke}.

In this regime, the relevant dynamics concern the quantum fluctuations of the cavity and mechanical modes around the classical steady-state described above. Rewriting each Heisenberg operator of Eqs.~(\ref{nonlinlang}) as the classical steady state value plus an additional
fluctuation operator with zero mean value, and neglecting all the nonlinear terms in the equations, one gets the following linearized QLE for the fluctuations
\begin{subequations}
\label{lle}
\begin{eqnarray}
\delta \dot{\hat{q}}& =&\Omega _{m}\delta \hat{p}, \label{lle1}\\
\delta \dot{\hat{p}}& =&-\left[\Omega _{m}+\partial_q^2 \omega(q_s)|\alpha_s|^2\right]\delta \hat{q}-\gamma _{m}\delta \hat{p}+G\delta \hat{X} \nonumber \\
&& +\hat{\xi} +\frac{\partial_q \kappa_1(q_s)\alpha_s}{\sqrt {\kappa_1(q_s)}}\hat{Y}_1^{in}, \label{lle2}\\
\delta \dot{\hat{X}}& =&-\left[\kappa_0+\kappa_1(q_s)\right] \delta \hat{X}+\Delta \delta \hat{Y}-\sqrt{2}\partial_q \kappa_1(q_s)\alpha_s \delta \hat{q} \nonumber \\
&& +
\sqrt{2\kappa_0}\hat{X}_0^{in}+\sqrt{2\kappa_1(q_s)}\hat{X}_1^{in}, \label{lle3}\\
\delta \dot{\hat{Y}}& =&-\left[\kappa_0+\kappa_1(q_s)\right] \delta \hat{Y}-\Delta \delta \hat{X}+G\delta \hat{q}\nonumber \\
&&+\sqrt{2\kappa_0}\hat{Y}_0^{in}+\sqrt{2\kappa_1(q_s)}\hat{Y}_1^{in}. \label{lle4}
\end{eqnarray}
\end{subequations}
These equations have been derived by choosing the phase reference of the cavity field so that $\alpha _{s}$ is real and positive, and by defining the detuning $\Delta = \omega(q_s)-\omega_L$. We have also introduced the cavity field
amplitude quadrature $\delta \hat{X} \equiv (\delta \hat{a}+\delta \hat{a}^{\dagger })/\sqrt{2}$, the phase quadrature $\delta \hat{Y} \equiv (\delta \hat{a}-\delta \hat{a}^{\dagger })/i\sqrt{2}$ and the
corresponding Hermitian input noise operators $\hat{X}_j^{in}\equiv (\hat{a}_j^{in}+\hat{a}_j^{in,\dag })/\sqrt{2}$ and $\hat{Y}_j^{in}\equiv (\hat{a}_j^{in}-\hat{a}_j^{in,\dag })/i\sqrt{2}$, $j=0,1$. The linearized QLE of Eqs.~(\ref{lle}) show that the mechanical and cavity mode
fluctuations are coupled by the effective optomechanical coupling
\begin{eqnarray}
G &=& -\partial_q \omega(q_s)\alpha_{s}\sqrt{2} \label{optoc} \\
&=& -2\partial_q \omega(q_s)\sqrt{\frac{\mathcal{P}\kappa_0 }{\hbar\omega _{L}\left[\left[\kappa_0+\kappa_1(q_s)\right]^{2}+\Delta ^{2}\right] }}, \nonumber
\end{eqnarray}
which can be made very large by increasing the intracavity amplitude $\alpha _{s}$.

The linearized QLE of Eqs.~(\ref{lle}) provide a complete description of the dynamical effects of optical absorption by the membrane. These QLE are identical to the set of QLE describing standard cavity optomechanical systems (see e.g. Refs.~\cite{amo,prl07,output}) except for some additional damping and noise terms: i) the cavity decay and noise terms related to $\kappa_1(q_s)$, describing the photon absorbed by the membrane and the vacuum optical field re-emitted within the cavity; ii) the noise term proportional to $Y_1^{in}$ describing heating of the mechanical resonator; iii) the dissipative coupling term $\sqrt{2}\partial_q \kappa_1(q_s)\alpha_s \delta q$. Actually, in Eq.~(\ref{lle2}) there is a further nonstandard term, given by $-\partial_q^2 \omega(q_s)|\alpha_s|^2\delta \hat{q}$, describing a shift of the mechanical frequency caused by the nonlinear dependence of $\omega(\hat{q})$ upon $\hat{q}$, which is not related with absorption. This latter term should be considered even when the quadratic term in $\omega(\hat{q})$ is very small, because it is proportional to the stationary number of intracavity photons $|\alpha_s|^2 $, which is necessarily very large in the linearized regime we are considering.

\section{Effect of membrane absorption on ground state cooling and stationary entanglement}

As discussed in detail in the literature \cite{amo,prl07,output,marquardt,wilson-rae,paternostro,wilson-rae2}, the radiation pressure interaction may lead a cavity optomechanical system into a stationary state with distinct quantum features, and specifically with the mechanical resonator very close to its quantum ground state, and showing robust continuous variable (CV) entanglement between the cavity and the vibrational modes. It is therefore interesting to verify if and when absorption by the membrane may hinder achieving the quantum regime for the cavity-membrane system under investigation.

Eqs.~(\ref{lle}) can be rewritten in compact form as $\dot{u}(t)=A u(t)+n(t)$, where $u^{T}(t)
=[\delta \hat{q}(t), \delta \hat{p}(t),\delta \hat{X}(t), \delta
\hat{Y}(t)]$ is the vector of fluctuation operators, $n^{T}(t) =[0,
\hat{\xi}(t) +\Gamma \hat{Y}_1^{in}(t)/\sqrt {2\kappa_1(q_s)},\sqrt{2\kappa_0}\hat{X}_0^{in}(t)+\sqrt{2\kappa_1(q_s)}\hat{X}_1^{in}(t), \sqrt{2\kappa_0}\hat{Y}_0^{in}(t)+\sqrt{2\kappa_1(q_s)}\hat{Y}_1^{in}(t)]$ is the noise vector, and the
matrix $A$ is the drift matrix
\begin{equation}\label{eq:drift}
A=\left(
\begin{array}{cccc}
0 & \Omega _{m} & 0 & 0 \\
-\Omega _{m} -h & -\gamma _{m} & G & 0 \\
-\Gamma & 0 & -\kappa_T(q_s)  & \Delta  \\
G & 0 & -\Delta  & -\kappa_T(q_s)
\end{array}%
\right) ,
\end{equation}%
where we have defined $\Gamma=\sqrt{2}\partial_q \kappa_1(q_s)\alpha_s$, and $h=\partial_q^2 \omega(q_s)|\alpha_s|^2$.
We are interested in the stationary state of the system, which is reached for $t \to \infty $ for every initial state if the system is stable (see Appendix B for the stability conditions, which we will consider to be always satisfied from now on). Due to the linearization and to the Gaussian nature of the noise operators, the steady state is a zero-mean Gaussian state, completely characterized by its covariance matrix (CM). The latter is given by the $
4\times 4$ matrix $V$ with elements $
V_{lm}=\left\langle u_{l}\left( \infty \right) u_{m}\left( \infty \right) +u_{m}\left( \infty \right) u_{l}\left( \infty \right)
\right\rangle /2$, where $u_{m}(\infty)$ is the asymptotic value of the $m$-th component of the vector $ u(t)$.
Using standard methods \cite{prl07,output}, one gets that the steady state CM $V$ is given by the solution of the Lyapunov equation
\begin{equation}
A V+V A^{T}=-D,  \label{Lyapunov}
\end{equation}%
with $A$ given by Eq.~(\ref{eq:drift}) and where $D$ is the $4\times 4$ diffusion matrix, which is related with the correlation functions of the noise vector elements by
$\left(\langle n_k(s)n_l(s')+ n_l(s')n_k(s)\rangle\right)/2 \equiv D_{kl} \delta(s-s')$, and whose explicit expression is
\begin{equation}\label{diffmat} \left(
\begin{array}{cccc}
0 & 0 & 0 & 0 \\
0 & \gamma _{m}\left( 2n_{0}+1\right) + \frac{\Gamma^2}{4\kappa_1(q_s)}& 0 & \frac{\Gamma}{2} \\
0 & 0 & \kappa_T(q_s)  & 0  \\
0 & \frac{\Gamma}{2} & 0  & \kappa_T(q_s)
\end{array}%
\right) .
\end{equation}%
Equation~(\ref{Lyapunov}) is a linear equation for the CM $V$ which can be solved in a straightforward way, but whose solution is very cumbersome and will not be reported here.
The CM contains all the information about the steady state: in particular, the mean energy of the mechanical resonator is given by
\begin{eqnarray} \label{meanenergy}
U &=& \frac{\hbar \Omega _{m}}{2}\left[ \left\langle \delta \hat{q}^{2}\right\rangle +\left\langle \delta \hat{p}^{2}\right\rangle \right]
= \frac{\hbar \Omega _{m}}{2}\left[ V_{11} +V_{22} \right] \\
&\equiv &\hbar \Omega _{m}\left( n+\frac{1}{2}\right),\nonumber
\end{eqnarray}
where $n$ is the effective mean vibrational number of the selected membrane mode. Obviously, in the
absence of radiation pressure coupling it is $n=n_{0}$, where $n_{0}$ corresponds to the actual temperature of the environment $T_{0}$. The
optomechanical coupling with the cavity mode can be used to engineer an effective bath of much lower temperature $T\ll T_{0}$, so that the
vibrational mode is cooled.
Moreover from the covariance matrix $V$ one can also calculate the entanglement between the cavity and the vibrational modes and at the steady state. We adopt as entanglement measure the logarithmic negativity
$E_{N}$, which is a convenient entanglement measure, easy to compute and also
additive. It is defined as \cite{logneg}
\begin{equation}
E_{N}=\max [0,-\ln 2\eta ^{-}], \label{eq:logneg}
\end{equation}
with
\begin{eqnarray}
&& \eta ^{-}\equiv 2^{-1/2}\left[ \Sigma (V)-\left[ \Sigma (V)^{2}-4\det V\right] ^{1/2}\right] ^{1/2}\\
&&\Sigma (V)\equiv \det V_{1}+\det V_{2}-2\det V_{c}, \label{lognegsigma}
\end{eqnarray}
where $V_{1},V_{2}$ and $V_{c}$ are the $2\times 2$ sub-block matrices of $V$
\begin{equation}
V\equiv \left(
\begin{array}{cc}
V_{1} & V_{c} \\
V_{c}^{T} & V_{2}%
\end{array}%
\right) .  \label{CMatrix}
\end{equation}
Determining $V$ from the Lyapunov equation of Eq.~(\ref{Lyapunov}) and using Eqs.~(\ref{meanenergy})-(\ref{lognegsigma}), one can therefore see if membrane absorption affects reaching a quantum regime for the cavity-membrane system.

At first sight, such absorption may be very harmful to quantum behavior, due to the additional damping and noise terms acting on both the optical and mechanical mode. In order to have an intuitive picture of how photon absorption by the membrane manifests its effects we focus on ground state cooling, i.e., on the evaluation of the steady state elements of the CM, $V_{11}$ and $V_{22}$ [see Eq.~(\ref{meanenergy})]. Following the approach of Refs.~\cite{output,Genes08}, one can equivalently determine the stationary CM, $V$, by solving the linearized QLE in the frequency domain rather than in the time domain. For the position and momentum variances of the mechanical resonator one gets (see Ref.~\cite{Genes08})
\begin{equation}
\left\langle \delta \hat{q}^{2}\right\rangle =\int_{-\infty }^{\infty }\frac{%
d\omega }{2\pi }S_{q}(\omega ),\;\;\;\;\left\langle \delta \hat{
p}^{2}\right\rangle =\int_{-\infty }^{\infty }\frac{d\omega }{2\pi }\frac{%
\omega ^{2}}{\Omega _{m}^{2}}S_{q}(\omega ),  \label{spectra}
\end{equation}%
where the position spectrum can be written as
\begin{equation} \label{posspe}
S_{q}(\omega )=\left|\chi _{\rm eff}(\omega )\right|^{2}[S_{th}(\omega )+S_{rp}(\omega)+S_{abs}(\omega)],
\end{equation}
with
\begin{equation}
S_{th}(\omega )=\frac{\gamma _{m}\omega }{\Omega _{m}}\coth \left( \frac{%
\hbar \omega }{2k_{B}T}\right) \simeq  \gamma _{m}\left( 2n_{0}+1\right)  \label{spectratherm}
\end{equation}%
the thermal noise spectrum,
\begin{equation}
S_{rp}(\omega)=\frac{G^{2}\kappa_T(q_s) \left[ \Delta ^{2}+\kappa_T^{2}(q_s)
+\omega ^{2}\right] }{\left[ \kappa_T^{2}(q_s)+(\omega -\Delta )^{2}\right] %
\left[ \kappa_T^{2}(q_s)+(\omega +\Delta )^{2}\right] }  \label{spectrarpn}
\end{equation}%
the radiation pressure noise spectrum, and finally
\begin{equation}
S_{abs}(\omega)=\frac{\Gamma^2}{4\kappa_1(q_s)}+\frac{\Gamma G \Delta \left[ \Delta ^{2}+\kappa_T^{2}(q_s)-\omega ^{2}\right] }{\left[ \kappa_T^{2}(q_s)+(\omega -\Delta )^{2}\right] %
\left[ \kappa_T^{2}(q_s)+(\omega +\Delta )^{2}\right] }  \label{spectraabs}
\end{equation}
the additional noise spectrum associated with membrane absorption, containing both the term quadratic in $\Gamma$ describing membrane heating, and the term linear in $\Gamma$, associated with the nonzero correlation between such heating noise and cavity phase noise [see also the diffusion matrix of Eq.~(\ref{diffmat})].

A further effect of optical absorption by the membrane is the modification of the effective mechanical susceptibility $ \chi _{\rm eff}(\omega )$ appearing in Eq.~(\ref{posspe}), which is given by
\begin{equation}
\chi _{\rm eff}(\omega )=\frac{\Omega _{m}}{\tilde{\Omega}_{m}^{2}-\omega^{2}-i\omega \gamma _{m}-\frac{G\Omega _{m}\left[G\Delta-\Gamma\left[\kappa_T(q_s) -i\omega\right]\right]}{\left[\kappa_T(q_s) -i\omega
\right]^{2}+\Delta ^{2}}} ,\label{chieffD}
\end{equation}%
where we have defined $\tilde{\Omega}_{m}^{2}=\Omega_{m}^{2}+h \Omega_{m}$.

The latter can be read as the susceptibility of an oscillator
with effective (frequency-dependent) resonance frequency and damping rate respectively given by
\begin{widetext}
\begin{eqnarray}
&&\Omega_{m}^{\rm eff}(\omega )=\left[ \tilde{\Omega} _{m}^{2}-\frac{G \Omega_{m}\left\{G \Delta \left[\kappa_T^{2}(q_s)-\omega ^{2}+\Delta ^{2}\right]-\Gamma\kappa_T(q_s)\left[\kappa_T^{2}(q_s)+\omega ^{2}+\Delta ^{2}\right]\right\}}{\left[ \kappa_T^{2}(q_s)+(\omega -\Delta )^{2}\right] %
\left[ \kappa_T^{2}(q_s)+(\omega +\Delta )^{2}\right] }\right]^{\frac{1}{2}},  \label{omegeff} \\
&&\gamma _{m}^{\rm eff}(\omega )=\gamma _{m}+\frac{G \Omega_{m}\left\{2G \Delta \kappa_T(q_s)-\Gamma\left[\kappa_T^{2}(q_s)+\omega ^{2}-\Delta ^{2}\right]\right\}}{\left[ \kappa_T^{2}(q_s)+(\omega -\Delta )^{2}\right] %
\left[ \kappa_T^{2}(q_s)+(\omega +\Delta )^{2}\right] }\label{dampeff}.
\end{eqnarray}%
\end{widetext}
As discussed in \cite{Dantan2008}, in the limiting case when the mechanical quality factor $Q_m$ is large enough, and the optomechanical coupling is not too strong, $G \ll \Omega_m $ (which is the relevant regime for ground state cooling), this effective susceptibility $ \chi _{\rm eff}(\omega )$ remains peaked at $\Omega_m$, and one can approximately calculate the variances of Eqs.~(\ref{spectra}) by evaluating the noise spectra just at the peak frequency $\Omega_m$. The integral are then easily performed and one gets
\begin{equation} \label{approx2}
n \simeq \left\langle \delta \hat{q}^{2}\right\rangle -\frac{1}{2} \simeq \left\langle \delta \hat{p}^{2}\right\rangle -\frac{1}{2}=
\frac{\gamma_m n_0 + \frac{\Gamma^2}{8\kappa_1(q_s)} +A_+}{\gamma_m + A_- - A_+},
\end{equation}
with
\begin{equation}\label{scattrate}
    A_{\pm}=\frac{G^2 \kappa_T + G \Gamma (\Delta \pm \Omega_m)}{\kappa_T^{2}(q_s)+(\Delta\pm \Omega_m )^{2}}.
\end{equation}
It is straightforward to verify that $A_{\pm}$ are just the scattering rates of laser photons into the antiStokes ($A_-$) and Stokes ($A_+$) sidebands centered at $\omega_L \mp \Omega_m$, respectively. Therefore $A_- - A_+$ is the net laser cooling rate and it is related to the effective mechanical damping by the relation $\gamma_m^{\rm eff}(\Omega_m)= \gamma_m+A_- - A_+$ \cite{Genes08}. The scattering rates of Eq.~(\ref{scattrate}) are the sum of two terms, corresponding to the scattering of photons through the cavity (the first term) or through membrane absorption (the second term). Eq.~(\ref{scattrate}) generalizes the expression of these rates obtained in Refs.~\cite{marquardt,wilson-rae,Genes08} to the case with non-negligible absorption. However the relevant result of the present analysis is that of Eq.~(\ref{approx2}), clearly showing that one can achieve ground state cooling when $A_- \gg A_+,\gamma_m$ and both $n_0$ and the term associated with membrane heating, $\Gamma^2/8\kappa_1(q_s)\gamma_m$, are not too large.

A simple description of the behavior of the logarithmic negativity is more difficult to find, but nonetheless one may reasonably expect that when the above conditions for achieving ground state cooling are met, one could also achieve stationary optomechanical entanglement as well. More generally speaking, the above analysis of ground state cooling suggests that it is still possible to find a parameter regime where quantum behavior of the stationary state of the membrane-in-the-middle setup is achievable, even in the presence of non-negligible absorption.

\begin{figure}
  \includegraphics[width=.45\textwidth]{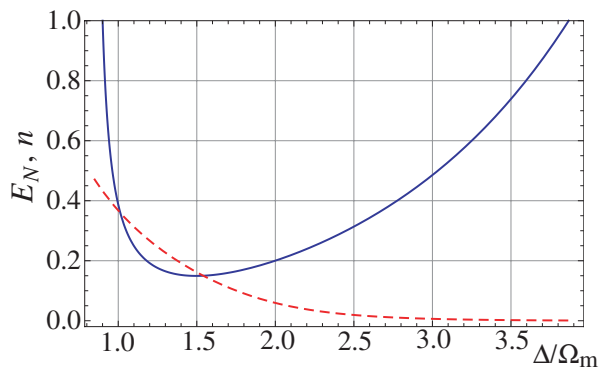}
  \caption{(Color online) Effective mean vibrational number $n$ (blue, full line) and logarithmic negativity  $E_{N}$ (red, dashed line) versus the scaled cavity detuning $\Delta/\Omega_m$, at fixed input power ${\mathcal P}=28.5$ mW and temperature $T_0=1$ K. Other parameters are: cavity length $L = 0.74$ mm, cavity finesse ${\mathcal F} = 23000$, membrane thickness $L_d= 50$ nm, membrane side-length $0.5$ mm, effective mode mass $m=8.5$ ng, mechanical resonance frequency $\Omega_m/2\pi=10$ MHz, and mechanical quality factor $Q_m=6 \times 10^6$.} \label{vsdetu}
\end{figure}

\begin{figure}
  \includegraphics[width=.45\textwidth]{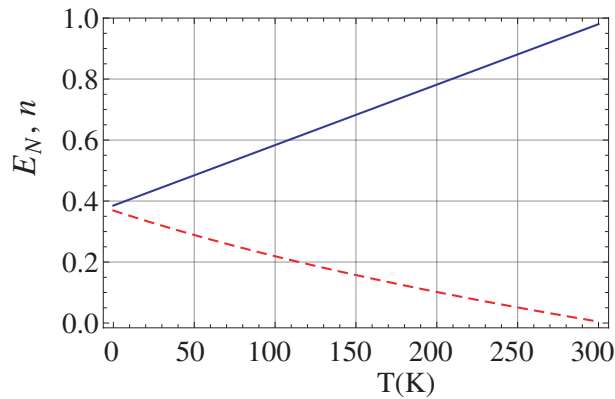}
  \caption{(Color online) Effective mean vibrational number $n$ (blue, full line) and logarithmic negativity  $E_{N}$ (red, dashed line) versus temperature at fixed input power ${\mathcal P}=28.5$ mW and detuning $\Delta=\Omega_m$. Remarkably, quantum features persist up to room temperature. The other parameters are as in Fig.~\protect\ref{vsdetu}.}\label{vstemp}
\end{figure}

\begin{figure}
 \includegraphics[width=.45\textwidth]{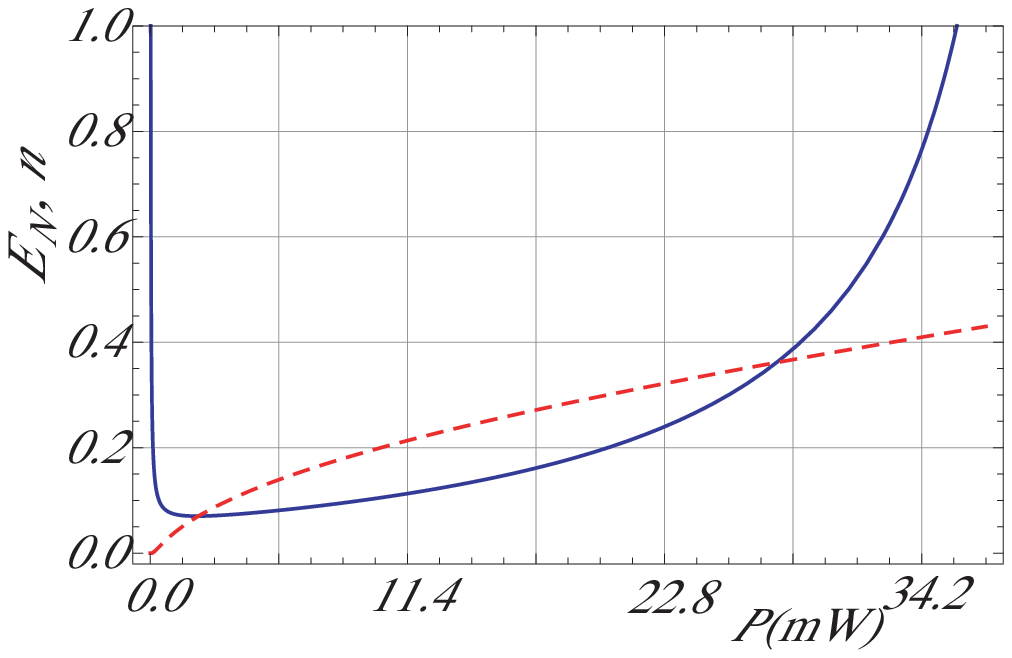}
  \caption{(Color online) Effective mean vibrational number $n$ (blue, full line) and logarithmic negativity  $E_{N}$ (red, dashed line) versus input power at fixed temperature $T_0=1$ K and detuning $\Delta=\Omega_m$. The other parameters are as in Fig.~\protect\ref{vsdetu}.} \label{vspower}
\end{figure}

\begin{figure}
   \includegraphics[width=.45\textwidth]{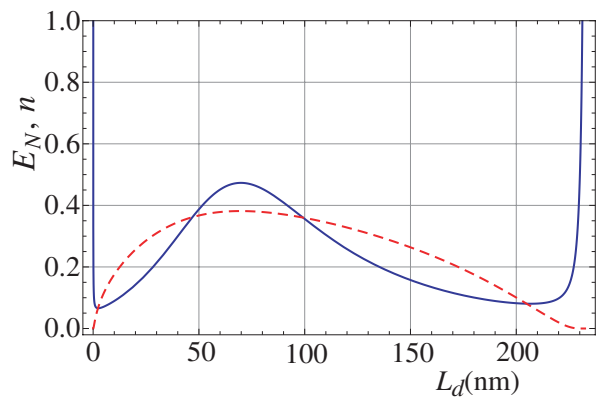}
  \caption{(Color online) Effective mean vibrational number $n$ (blue, full line) and logarithmic negativity  $E_{N}$ (red, dashed line) versus membrane thickness at fixed temperature $T_0=1$ K, input power ${\mathcal P}=28.5$ mW, and detuning $\Delta=\Omega_m$. The other parameters are as in Fig.~\protect\ref{vsdetu}.}
  \label{vsthick}
\end{figure}

We have verified these predictions by considering parameters similar to the experiment of Ref.~\cite{kimble2}. We have verified in particular that the approximate expression of Eq.~(\ref{approx2}) satisfactorily describes the vibrational occupancy $n$ at not too large input power. To be more specific we have taken
a SiN membrane with refraction index $n_M = 2.0 + \imath 10^{-5}$, side length $0.5$ mm and thickness $L_d=50$ nm, implying an effective mass $m = 8.5$ ng; we also assume a mechanical resonance frequency $\Omega_m/2\pi$ = 10 MHz, a mechanical quality factor $Q_m = 6 \times 10^6$, and a cavity with length $L=0.74$ mm and finesse ${\mathcal F} = 23000$. The results are shown in Figs.~\ref{vsdetu}-\ref{vsthick}, where we plot simultaneously the mean vibrational occupancy $n$ and the logarithmic negativity $E_N$ as a function of various parameters. All figures show that despite absorption may appreciably affect the cavity finesse, one can choose an operation point where the cavity-membrane system can be put in a steady state with distinct quantum features, i.e., with a vibrational mode close to its ground state and a significant optomechanical entanglement.

In Fig.~\ref{vsdetu} $n$ and $E_N$ are plotted versus the cavity detuning $\Delta$ (at fixed reservoir temperature $T_0=1$ K and laser input power ${\mathcal P}=28.5$ mW) and one can see that, as known in the literature \cite{prl07,marquardt,wilson-rae,output}, one has significative simultaneous cooling and entanglement when $\Delta \sim \Omega_m$, i.e., when the cavity is resonant with the anti-Stokes sideband of the driving laser. Fig.~\ref{vstemp} shows $n$ and $E_N$ versus temperature $T_0$ (at fixed detuning $\Delta = \Omega_m$ input power ${\mathcal P}=28.5$ mW) and, as expected, they both worsen for increasing temperature; it is remarkable however that quantum features $n<1$ and $E_N>0$ persists up to room temperatures, $T_0 \simeq 300$ K in this optimal parameter choice. In Fig.~\ref{vspower} instead we keep $\Delta = \Omega_m$ and $T_0=1$ K fixed and we plot $n$ and $E_N$ versus the input power ${\mathcal P}$. Increasing ${\mathcal P}$ means increasing the effective optomechanical coupling $G$ and therefore also optomechanical entanglement increases; on the contrary $n$ is minimum at not too large input power and therefore at not too large coupling $G$, because for larger $G$ higher-order scattering process tend to heat the vibrational motion (see Refs.~\cite{output,marquardt,wilson-rae,wilson-rae2}). Finally in Fig.~\ref{vsthick} we study the dependence of $n$ and $E_N$ upon the membrane thickness $L_d$, by keeping all the other parameter fixed ($\Delta = \Omega_m$, $T_0=1$ K, ${\mathcal P}=28.5$ mW). We notice that there are two interesting parameter regions showing good simultaneous cooling and entanglement: one at small thickness $L_d \sim 30$ nm, and a second one at larger thickness, $L_d \simeq 150$ nm. Small membrane thickness is obviously advantageous for quantum effects because it minimizes both mass $m$ and optical absorption effects; however one has appreciable cooling and entanglement also for a thicker membrane because in this case, even if the mass $m$ and absorption are larger, one has maximum membrane reflectivity ${\mathcal R}(k_0)$. In fact Eq.~(\ref{eq:refl}) shows that ${\mathcal R}(k_0)$ is maximum when $n_M k_0 L_d =\pi/2$, i.e. for $L_d = \lambda/4n_M$, which corresponds to $L_d \simeq 130$ nm for a laser wavelength $\lambda=1064$ nm.

\section{Conclusions}

We have analyzed the cavity optomechanical system formed by a Fabry-Perot cavity with a thin semi-transparent membrane placed at its center. In particular we have investigated the effect of optical absorption by the membrane, which is usually neglected in standard treatments. Membrane absorption has an appreciable effect on the cavity finesse and we have studied if and when it may also hinder achieving a quantum stationary state of the cavity-membrane system. Such absorption is responsible for a modification of the scattering rates of laser photons into the Stokes and antiStokes sidebands, and for an effective increase of the initial membrane temperature due to absorption-induced heating. Considering state-of-the-art experimental parameters, we have seen that one can always optimize the operation point so that membrane absorption does not pose serious limitations for reaching quantum effects. Remarkably, using parameter analogous to those of Ref.~\cite{kimble2}, ground state cooling and stationary optmechanical entanglement may persist up to room temperatures.
The fact that stationary optomechanical entanglement is possible even with non-negligible absorption is particularly relevant: appreciable optomechanical
entanglement is achieved only at large enough intracavity power: in this regime however membrane heating by absorption becomes relevant and it is not
obvious that optomechanical entanglement is robust even against such heating. Stationary entanglement is particularly important because it lasts for ever and can be extremely useful for storing and redistributing quantum information, or realizing quantum interfaces \cite{fam,atom-membrane}. Finally, the membrane-in-the-middle setup could be exploited also for quantum limited detection of weak forces \cite{ligocooling}.

\emph{Note added in proof} Recently Ref.~\cite{law2} derived the multimode Hamiltonian of the one-dimensional limit of the membrane-in-the-middle scheme, without making the linear approximation in the interaction. The multimode Hamiltonian derived here in Appendix A coincides with that of Ref.~\cite{law2} when the one-dimensional and linearized limit is taken.

\section{Acknowledgments}

This work has been supported by the European Commission (FP-7 FET-Open project MINOS), and by INFN (SQUALO project).

\begin{appendix}

\section{General multi-mode Hamiltonian of the system} \label{bigappen}

An empty Fabry-Perot cavity supports an infinite set of optical modes, well described by the Hermite-Gauss modes \cite{siegman}.
In the case of a symmetric cavity formed by two identical spherical mirrors with radius of curvature $R$, separated by a distance $L$, in the coordinate system with the $z$ axis along the cavity axis and centered at the cavity center (see Fig.~1), these Hermite-Gauss modes are written as
\begin{equation}\label{eq:herm-gau0}
u_{m n p}^0(x,y,z)= T_{m n p}(x,y,z)\sin\left[\Phi_{m n p}(x,y,z)-\frac{p \pi}{2}\right],
\end{equation}
where, for $m,n=0,1\ldots$, $p=1,2,\ldots$,
\begin{equation}
\label{eq:herm-gau1}
T_{m n p}(x,y,z) = \frac{H_n\left[\frac{\sqrt{2}x}{w(z)}\right]H_m\left[\frac{\sqrt{2}y}{w(z)}\right]\exp\left[-\frac{x^2+y^2}{w^2(z)}\right]}{w(z)\sqrt{\pi 2^{n+m-2}n! m! L }},
\end{equation}
\begin{equation}
\Phi_{m n p}(x,y,z) = k z - \psi(z)(m+n+1)+k\frac{x^2+y^2}{2 R(z)} \label{eq:herm-gau2}.
\end{equation}
Here $H_n(x)$ is the $n$-th Hermite polynomial, $w(z)= w_0\sqrt{1+z^2/z_R^2}$ is the width of the cavity mode at $z$, $R(z)=z+z_R^2/z$ is the radius of curvature of the wavefront at $z$, $\psi(z)=\arctan (z/z_R)$ is the Gouy phase shift \cite{siegman}. The cavity waist $w_0$ is related to the cavity geometry by the relation $w_0^2= (L/k)\sqrt{(1+g)/(1-g)}$, where $g=1-L/R$, and the Rayleigh range $z_R$ is given by $z_R=k w_0^2/2$. Each mode oscillates in time with an angular frequency $\omega_{m n p}^0$ given by
\begin{equation}\label{eq:omemnp}
    \omega_{m n p}^0=\frac{c \pi}{L}\left[p+\frac{m+n+1}{\pi}\arccos g\right] ,
\end{equation}
which is determined by the boundary condition $\Phi_{m n p}(0,0,L/2)=-\Phi_{m n p}(0,0,-L/2)=p\pi/2$. The mode functions $u_{m n p}^0(x,y,z)$ are proportional to the electric field and are normalized according to
$$
\int dx \int dy \int dz u_{m n p}^0(x,y,z) u_{m' n' p'}^0(x,y,z)= \delta_{m m'} \delta_{n n'} \delta_{p p'}.
$$
Notice that the mode functions depends upon $m$, $n$, and $p$ also implicitly, through the dependence of the wave vector $k=\omega_{m n p}/c$.

\begin{figure}[h]
   \centering
   \includegraphics[width=.48\textwidth]{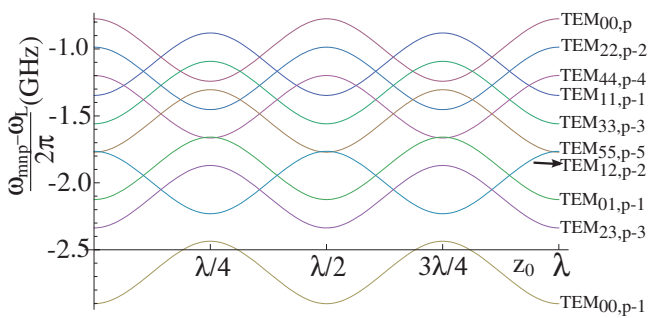}
   \caption{(Color online) Frequency shift with respect to a driving laser with wavelength $\lambda = 1064$ nm of some of the Hermite-Gauss modes versus the membrane position along the cavity axis $z_0$. The plot refers to a cavity with length $L=9$ cm, mirror radius of curvature $R=10$ cm, and to a membrane with thickness $L_d = 50$ nm, and index of refraction $n_M =2.0$, which corresponds to a reflectivity $\sqrt{{\mathcal R}} \simeq 0.4$. }
   \label{spectro}
\end{figure}

The thin membrane is a dielectric slab of thickness $L_d$ and index of refraction $n_M$, and when it is placed within the cavity the mode functions and their frequency change in a way which is dependent upon the position and orientation of the membrane with respect to the cavity. In order to minimize diffraction losses, the membrane should be placed orthogonal to the cavity axis and exactly at the cavity waist, so that the optical wavefront coincides as much as possible with the membrane surface. In the limiting case of a thin membrane, placed at $z_0$ close to the waist, well within the Rayleigh range so that  $z_0 \pm L_d/2 \ll z_R$ (see Fig.~1), the above Hermite-Gauss modes can be still used to describe the modified optical modes. In fact in this limit the new mode functions are well approximated by
\begin{widetext}
\begin{equation}\label{eq:modes-membr}
    u_{m n p}(x,y,z)=\left\{\begin{array}{cc}
                              A_{m n p} T_{m n p}(x,y,z)\sin\left[\Phi_{m n p}(x,y,z)+\Phi_{m n p}(0,0,L/2)\right] & -L/2 < z < z_0-L_d/2 \\
                             C_{m n p} T_{m n p}(x,y,z)\sin\left[\Phi_{m n p}(x,y,z)+\varphi_{m n p} \right] & z_0-L_d/2 < z < z_0+L_d/2 \\
                              B_{m n p} T_{m n p}(x,y,z)\sin\left[\Phi_{m n p}(x,y,z)-\Phi_{m n p}(0,0,L/2)\right] & z_0+L_d/2 < z < L/2,
                            \end{array} \right.
\end{equation}
\end{widetext}
so that they automatically satisfies the null boundary condition at the mirrors. The dimensionless parameters $A_{m n p}$, $B_{m n p}$, $C_{m n p}$, and $\varphi_{m n p}$ are determined by matching the solutions at the two membrane boundaries. The boundary conditions also determine the implicit equation for the new mode frequencies $\omega_{m n p} = ck_{m n p}$,
\begin{eqnarray}
&& \sin\left[k_{m n p}L-2(n+m+1)\psi(L/2)+\beta(k_{m n p})\right] \nonumber \\
&&=\sqrt{{\mathcal R}(k_{m n p})}\cos(2k_{m n p}z_0), \label{eq:freq-membr-exact}
\end{eqnarray}
where
\begin{equation}
{\mathcal R}(k)=\frac{(n_M^2-1)^2}{4n_M^2\cot^2 (n_M k L_d)+(n_M^2+1)^2} \label{eq:refl}
\end{equation}
is the membrane intensity reflection coefficient \cite{Broker}, and
\begin{eqnarray}
&& \beta(k) = \arccos\left[\frac{2n_M \cot (n_M k L_d)}{\sqrt{4n_M^2\cot^2 (n_M k L_d)+(n_M^2+1)^2}}\right] \nonumber \\
&&-kL_d =\arcsin \left[\frac{n_M^2+1}{n_M^2-1}\sqrt{{\mathcal R(k)}}\right]-kL_d .
\end{eqnarray}
The membrane causes a frequency shift of each cavity mode which depends upon the position of the membrane along the cavity axis $z_0$. However, this shift is always much smaller than the bare frequency, i.e., the solution of Eq.~(\ref{eq:freq-membr-exact}) can be written as $k_{m n p}=k_{m n p}^0+\delta k_{m n p}$, with $|\delta k_{m n p} | \ll k_{m n p}^0$, and using the fact that
$\Phi_0(k_{m n p}^0)\equiv 2\Phi_{mnp}(0,0,L/2)=k_{m n p}^0 L - 2(n+m+1)\psi(L/2)=p \pi$,
one can derive the explicit expression of the frequency shift of a given cavity mode with unperturbed frequency $\omega_{n m p}^0=c k_{n m p}^0$
\begin{widetext}
\begin{equation}\label{eq:freq-membr-shift}
    \delta \omega_{n m p}=\frac{c}{L}\left\{(-1)^{\left[\frac{n_M k_{n m p}^0 L_d}{\pi}\right]}\arcsin
    \left[(-1)^p\sqrt{{\mathcal R\left(k_{n m p}^0\right)}}\cos(2k_{n m p}^0z_0)\right]-\beta(k_{n m p}^0)-\left[\frac{n_M k_{n m p}^0 L_d}{\pi}\right]\pi \right\},
\end{equation}
\end{widetext}
with $[x]$ denoting the integer part of $x$. A selection of the modified cavity mode frequencies $\omega_{m n p}=\omega_{m n p}^0+\delta \omega_{n m p}$ in the presence of the membrane are plotted versus the membrane position along the cavity axis $z_0$ in Fig.~\ref{spectro}. As predicted by Eq.~(\ref{eq:freq-membr-shift}) they are periodic with a period equal to half wavelength.

The optical modes are therefore described by the ``free optical'' Hamiltonian
\begin{equation}\label{eq:Ham-opt}
    H_{opt}=\sum_l \hbar \omega_l \hat{a}_l^{\dagger} \hat{a}_l,
\end{equation}
where $\omega_l=\omega_{m n p}$ are the frequencies of the optical modes in the presence of the membrane and $l \leftrightarrow \{m n p\}$ is a collective index. The creation operator $\hat{a}_l^{\dagger}$ creates a photon in the cavity mode with the mode function $u_{m n p}(x,y,z)$ of Eq.~(\ref{eq:modes-membr}), and satisfies the boson commutation rule $\left[\hat{a}_l,\hat{a}_{l '}^{\dagger}\right]=\delta_{l l '}$.

Considering the membrane motion means assuming that its position along the cavity axis oscillates in time, $z_0 \to z_0+z(x,y,t)$, where $z(x,y,t)$ gives the membrane transverse deformation field. One typically assumes the high stress regime of a taut membrane, in which bending effects are negligible
and one can use the classical wave equation to describe the membrane vibration. In such a limit the (normalized) vibrational normal modes of the square membrane with length $D$ are
\begin{equation}\label{eq:vib-modes}
    \phi_{j k}(x,y)=\frac{2}{D}\sin\left(\frac{j\pi x}{D}+\frac{j \pi}{2}\right)\sin\left(\frac{k\pi y}{D}+\frac{k \pi}{2}\right),\;\;
\end{equation}
with $j,k=1,2,\ldots$, $|x|,|y| \leq D/2$. The vibrational mode with $j,k$ indices oscillates in time with eigenfrequency $\Omega_{j k}=(c_s \pi/D)\sqrt{j^2+k^2}$, where $c_s=\sqrt{T/\sigma}$ is the sound velocity, with $T$ the surface tension and $\sigma$ the surface mass density. The membrane motion is then quantized by associating to each normal mode a bosonic annihilation operator $\hat{b}_{i}$ ($i \leftrightarrow \{jk\}$ is the membrane collective index) such that $\left[\hat{b}_{i},\hat{b}_{i'}^{\dagger}\right]=\delta_{i i'}$. The free vibrational Hamiltonian can then be written as
\begin{equation}\label{eq:Ham-vib}
    H_{vib}=\sum_i \hbar \Omega_i \hat{b}_i^{\dagger} \hat{b}_i = \sum_i \frac{\hbar \Omega_i}{2} \left(\hat{p}_i^2+\hat{q}_i^2\right),
\end{equation}
where we have introduced the dimensionless position $\hat{q}_i=(\hat{b}_i^{\dagger} +\hat{b}_i)/\sqrt{2}$ and momentum $\hat{p}_i=(\hat{b}_i-\hat{b}_i^{\dagger})/\sqrt{2}$ operators of each vibrational normal mode. The membrane deformation field operator can then be written in the equivalent forms
\begin{eqnarray}\label{eq:vib-field}
    && z(x,y)=\sum_i\sqrt{\frac{\hbar}{2 \sigma \Omega_i}}\phi_{i}(x,y)(\hat{b}_i^{\dagger} +\hat{b}_i)  \\
    &&=
    \sum_i\sqrt{\frac{\hbar}{m \Omega_i}}\hat{q}_i\sin\left(\frac{j\pi x}{D}+\frac{j \pi}{2}\right)\sin\left(\frac{k\pi y}{D}+\frac{k \pi}{2}\right), \nonumber
\end{eqnarray}
where $m=\sigma D^2/4$ is the effective mass of each vibrational mode.

\subsection{The radiation pressure interaction Hamiltonian}

At lowest order in the membrane deformation field $z(x,y)$, one can write the interaction Hamiltonian (in the Schr\"odinger picture) between the optical field in the cavity and the membrane motion as \cite{pinard}
\begin{equation}\label{eq:hint-radpress}
   H_{int}=-\int_{-D/2}^{D/2}dx \int_{-D/2}^{D/2}dy P(x,y) z(x,y),
\end{equation}
where $P(x,y)$ is the radiation pressure exerted by the cavity field on the membrane. A direct way of evaluating this radiation pressure is to start from the Lorenz force per unit volume $\vec{f}(x,y,z)=\rho(x,y,z)\vec{E}(x,y,z)+\vec{J}(x,y,z)\times \vec{B}(x,y,z)$ within the membrane \cite{loudon} and integrate it over the membrane length, that is
\begin{equation}\label{eq:radpress}
   P(x,y)=\int_{z_0-L_d/2}^{z_0+L_d/2}dz f_z(x,y,z).
\end{equation}
The electric field $\vec{E}(x,y,z)$ is polarized parallel to the membrane and therefore its force contribution is balanced by the membrane support; moreover the current density ${\vec J}$ is due only to bound charges, $\vec{J}=\partial \vec{P}/\partial t$ with ${\vec P}$ the polarization vector, and therefore the radiation pressure acting on the membrane can be written as
\begin{equation}\label{eq:hint-radpress-lorenz}
   P(x,y)=\varepsilon_0 (n_M^2-1)\int_{z_0-\frac{L_d}{2}}^{z_0+\frac{L_d}{2}}dz \left[\frac{\partial \vec{E}}{\partial t}(x,y,z) \times \vec{B}(x,y,z)\right]_z.
\end{equation}
The $\hat{x}$ axis can be taken along the direction of field polarization and therefore the electric field operator within the cavity can be written (in the interaction picture) as
\begin{equation}\label{eq:efield-cav}
   \vec{E}(x,y,z,t)=\hat{x}\sum_l\sqrt{\frac{\hbar \omega_l}{2\varepsilon_0}} u_l(x,y,z)i \left(\hat{a}_l e^{-i\omega_l t}-\hat{a}_l^{\dagger} e^{i\omega_l t}\right),
\end{equation}
where $u_l(x,y,z)$ are the normalized mode functions specified by Eqs.~(\ref{eq:modes-membr}) and we have used the collective index $l \leftrightarrow \{m n p\}$. The membrane is placed close to the waist, well within the Rayleigh range $z_0\pm \L_d/2 \ll z_R$, and this allows to simplify considerably $u_l(x,y,z)$ and therefore $\vec{E}(x,y,z,t)$ within the membrane. In fact, one can take $w(z) \simeq w_0$, $\psi(z) \simeq 0$, and $R(z) \simeq \infty$ in Eq.~(\ref{eq:herm-gau2}), so that $\Phi_l(x,y,z)\simeq n_M k_l z$, and by using Eq.~(\ref{eq:modes-membr}), one can rewrite Eq.~(\ref{eq:efield-cav}) within the membrane as
\begin{widetext}
\begin{equation}
 \vec{E}(x,y,z,t)=\hat{x}\sum_l\sqrt{\frac{\hbar \omega_l}{\varepsilon_0}} \frac{C_l\tilde{T}_l(x,y)}{\sqrt{L}}\sin(n_M k_l z+\varphi_l)  i \left(\hat{a}_l e^{-i\omega_l t}-\hat{a}_l^{\dagger} e^{i\omega_l t}\right), \label{eq:efield-mem}
\end{equation}
where, from Eq.~(\ref{eq:herm-gau1}),
\begin{equation}\label{eq:herm-gau1-simpl}
    \tilde{T}_{l}(x,y) = \frac{H_n[\sqrt{2}x/w_0]H_m[\sqrt{2}y/w_0]}{ w_0\sqrt{\pi 2^{n+m-1}n! m!}}\exp\left\{-\frac{x^2+y^2}{w_0^2}\right\}
\end{equation}
is the normalized transverse pattern of the optical field within the membrane. The corresponding expression of the magnetic field within the membrane is given by
\begin{equation}\label{eq:bfield-mem}
\vec{B}(x,y,z)=\hat{y}\frac{n_M}{c}\sum_l\sqrt{\frac{\hbar \omega_l}{\varepsilon_0}} \frac{C_l\tilde{T}_l(x,y)}{\sqrt{L}} \cos(n_M k_l z+\varphi_l) \left(\hat{a}_l e^{-i\omega_l t}+\hat{a}_l^{\dagger} e^{i\omega_l t}\right),
\end{equation}
so that, inserting Eqs.~(\ref{eq:efield-mem}) and (\ref{eq:bfield-mem}) into Eq.~(\ref{eq:hint-radpress-lorenz}) and performing the integration, one gets the following general expression for the radiation pressure operator of the cavity field on the membrane (in the Schr\"odinger picture)
\begin{equation}\label{eq:radpress-gen-app}
   P(x,y)=\frac{(n_M^2-1)\hbar}{L}\sum_{l,m}\sqrt{\omega_l \omega_{m}} \tilde{T}_l(x,y)\tilde{T}_m(x,y)\tilde{\Lambda}_{l,m}\left(\hat{a}_l \hat{a}_m+\hat{a}_l \hat{a}_m^{\dagger} +\hat{a}_l^{\dagger} \hat{a}_m+\hat{a}_l^{\dagger}\hat{a}_m^{\dagger}\right),
\end{equation}
with the dimensionless ``longitudinal'' factor $\tilde{\Lambda}_{l,m}$
\begin{equation}\label{eq:radpress-coeff}
\tilde{\Lambda}_{l,m}=C_l C_m\frac{k_l}{k_l+k_m}\sin\left[\frac{n_M (k_l+k_m)L_d}{2}\right]\sin[n_M (k_l+k_m)z_0+(\varphi_l+\varphi_m)].
\end{equation}
This factor depends upon the constants $C_l$ and $\varphi_l$ associated with the cavity mode functions within the membrane (see Eq.~(\ref{eq:modes-membr})). Notice that a similar term in $\tilde{\Lambda}_{l,m}$ with $k_l+k_m \rightarrow k_l-k_m$ gives no contribution when inserted into the sum of Eq.~(\ref{eq:radpress-gen-app}) because it is antisymmetric with respect to $l \leftrightarrow m$ exchanges. The cavity mode constants $C_l$ and $\varphi_l$ depends upon the physical parameters of the cavity-membrane system, and their explicit expression is determined by the boundary conditions and by the normalization. After lengthy but straightforward calculations one gets
\begin{eqnarray}\label{eq:radpress-coeff-fin}
&& \Lambda_{l,m}\equiv \left(n_M^2-1\right) \tilde{\Lambda}_{l,m} = \frac{k_l}{k_l+k_m}\sin\left[\frac{n_M (k_l+k_m)L_d}{2}\right]\frac{\sin(2k_m z_0)}{\sin(n_M k_l L_d)}\left[\frac{\sin\left[\Phi_0(k_l)-k_l L_d\right]}{\sin\left[\Phi_0(k_m)-k_m L_d\right]}\frac{1+\sqrt{1-s(k_l)^2}}{1+\sqrt{1-s(k_m)^2}}\right]^{1/2} \nonumber \\
&&\times \left[\frac{{\mathcal R(k_l)}{\mathcal R(k_m)}}{\left[1-{\mathcal R(k_m)}\cos^2(2k_m z_0)\right]\left[1-{\mathcal R(k_l)}\cos^2(2k_l z_0)\right]}\right]^{1/4} \left[1+\frac{s(k_l)}{s(k_m)}\frac{1+\sqrt{1-s(k_m)^2}}{1+\sqrt{1-s(k_l)^2}}\right],
\end{eqnarray}
with ${\mathcal R(k)}$ the reflectivity given by Eq.~(\ref{eq:refl}), and
\begin{eqnarray}
&&\Phi_0(k)=2\Phi_{mnp}(0,0,L/2)=kL-2(n+m+1)\arctan\left(\frac{L}{2z_R}\right), \\
&& s(k)=\frac{\sin(2k z_0)\sin(n_M k L_d)}{\sin\left[\Phi_0(k)-k L_d\right]}. \label{eq:essekap}
\end{eqnarray}
\end{widetext}
It is easy to verify from Eq.~(\ref{eq:radpress-coeff-fin}) that when $j=l$ the longitudinal factor $\Lambda_{j,l}$ becomes much simpler
\begin{equation}\label{eq:radpress-coeff-fin-diag}
\Lambda_{l,l}=\sin(2k_l z_0)\sqrt{\frac{{\mathcal R(k_l)}}{1-{\mathcal R(k_l)}\cos^2(2k_l z_0)}}.
\end{equation}
By comparing with Eq.~(\ref{eq:freq-membr-shift}), one can see that $\Lambda_{l,l} \propto \partial \omega_l/\partial z_0$, as it must be expected from the fact that the interaction Hamiltonian of Eq.~(\ref{eq:hint-radpress}) is at first order in the membrane deformation field $z(x,y,t)$.
Eq.~(\ref{eq:radpress-coeff-fin-diag}) is often useful in practice because one can safely approximate $\Lambda_{j,l} \simeq \Lambda_{l,l}$ in the expression of the radiation pressure field of Eq.~(\ref{eq:radpress-gen-app}) whenever $|k_j-k_l|\ll k_l$, which is verified when the driving laser is not too broad, so that only the cavity modes within a narrow frequency band are significantly populated.

Eqs.~(\ref{eq:radpress-gen-app})-(\ref{eq:essekap}) provide the general expression of the radiation pressure operator acting on the membrane modes which, inserted within Eq.~(\ref{eq:hint-radpress}) yields the optomechanical interaction Hamiltonian at first order in the membrane deformation field $z(x,y)$. Using the explicit expression of Eq.~(\ref{eq:vib-field}), one finally gets the general expression of the radiation pressure optomechanical interaction for the membrane-in-the-middle configuration
\begin{eqnarray}
   H_{int}&=& -\frac{\hbar}{L}\sum_{i,j,l} \sqrt{\frac{\hbar\omega_j \omega_l}{m \Omega_i}}\Theta_{i,j,l}\Lambda_{j,l}\hat{q}_i \left(\hat{a}_j \hat{a}_l+\hat{a}_j \hat{a}_l^{\dagger}\right. \nonumber \\
   && \left. +\hat{a}_j^{\dagger} \hat{a}_l+\hat{a}_j^{\dagger}\hat{a}_l^{\dagger}\right), \label{eq:hint-radpress-expl}
\end{eqnarray}
where we have defined the dimensionless \emph{transverse overlap integral} $\Theta_{i,j,l}$ between the $j$-th and $l$-th cavity modes and the $i$-th membrane vibrational mode,
\begin{equation}\label{eq:overlap}
    \Theta_{i,j,l}=\frac{D}{2}\int_{-D/2}^{D/2}dx \int_{-D/2}^{D/2}dy \phi_i(x,y) \tilde{T}_j(x,y)\tilde{T}_l(x,y) .
\end{equation}
Due to the normalization of the involved mode functions, these overlaps satisfy $ |\Theta_{i,j,l}| \leq 1$.

Eq.~(\ref{eq:hint-radpress-expl}) generalizes in various directions existing treatments of the radiation pressure interaction Hamiltonian. In fact, it is possible to verify that it reproduces the radiation pressure interaction Hamiltonian of Ref.~\cite{law} if one restricts to the longitudinal axis only and considers a one-dimensional cavity with a moving perfectly reflecting mirror ${\mathcal R}(k)=1$. It is also straightforward to verify that Eq.~(\ref{eq:hint-radpress-expl}) reproduces the expression of the radiation pressure operator of Ref.~\cite{loudon} in the \emph{free-space} limit when there is no cavity.

Since we are considering \emph{driven optical} systems, the general interaction Hamiltonian of Eq.~(\ref{eq:hint-radpress-expl}) can be further simplified. In fact, at optical frequencies the counter-rotating terms $\hat{a}_j \hat{a}_l+ h.c$ have always negligible effects, and this holds true for the vacuum contribution associated with the Casimir force, as well \cite{loudon}. As a consequence, for a typical membrane-in-the-middle system the multi-mode interaction Hamiltonian simplifies to
\begin{equation}\label{eq:hint-radpress-pract}
    H_{int}=-\frac{2\hbar}{L}\sum_{i,j,l} \sqrt{\frac{\hbar\omega_j \omega_l}{m \Omega_i}}\Theta_{i,j,l}\Lambda_{j,l}\hat{q}_i \hat{a}_j^{\dagger} \hat{a}_l.
\end{equation}
Therefore, the total Hamiltonian of the cavity-membrane system in the absence of driving is
\begin{equation}\label{eq:htot}
    H_{tot}=H_{opt}+H_{vib}+H_{int}
\end{equation}
with $H_{opt}$ given by Eq.~(\ref{eq:Ham-opt}), $H_{vib}$ given by Eq.~(\ref{eq:Ham-vib}), and $H_{int}$ by Eq.~(\ref{eq:hint-radpress-pract}).

Eq.~(\ref{eq:hint-radpress-pract}) shows that radiation pressure leads in general to a trilinear interaction describing all possible scattering processes in which a photon from one cavity mode is scattered to another one, mediated by the emission or absorption of a vibrational phonon. The optomechanical coupling rates associated with these scattering processes depends upon both cavity and membrane properties, as well as the membrane position, as illustrated by Eq.~(\ref{eq:hint-radpress-pract}) together with Eqs.~(\ref{eq:radpress-coeff-fin}) and (\ref{eq:overlap}).

\subsection{Reduction to the single-mode-single-resonator case}

One can restrict to a single cavity mode whenever the driving laser is tuned and aligned in order to be matched and to populate a given cavity mode only (here associated with the annihilation operator $\hat{a}$), and if scattering from the driven mode to other modes is negligible. Photon scattering by the membrane can be neglected provided that the membrane is not misaligned and not too far from the cavity waist, and if the motional sidebands of the driven mode do not overlap with nearby cavity modes. This latter condition is satisfied when the relevant mechanical frequencies $\Omega_i$ are smaller than the typical cavity mode separation, which is of the order of the free spectral range $c/2L$. Under these conditions, the interaction Hamiltonian of Eq.~(\ref{eq:hint-radpress-pract}) reduces to
\begin{equation}\label{eq:hint-interm}
    H_{int}=-\hat{a}^{\dagger} \hat{a}\sum_{i} \hbar g_i\hat{q}_i ,
\end{equation}
where $g_i$ is the vacuum optomechanical coupling rate between the selected cavity mode and the $i$-th vibrational mode of the membrane, given by
\begin{equation}\label{eq:gvacuumi}
    g_i=\frac{2\hbar \omega_0}{L}\sqrt{\frac{\hbar}{m \Omega_i}}\Theta_{i,0,0}\Lambda_{0},
\end{equation}
with $\omega_0 = c k_0$ the frequency of the driven cavity mode, $\Theta_{i,0,0}$ the transverse overlap of Eq.~(\ref{eq:overlap}), and $\Lambda_0=
\sin(2k_0 z_0)\left\{{\mathcal R(k_0)}/\left[1-{\mathcal R(k_0)}\cos^2(2k_0 z_0)\right]\right\}^{1/2}$ [see Eq.~(\ref{eq:radpress-coeff-fin-diag})]. Eq.~(\ref{eq:hint-interm}) shows that the driven cavity mode is directly coupled to a
collective membrane vibrational operator $ \hat{q}_{eff} \propto \sum_i g_i\hat{q}_i$. However, if the detection bandwidth is chosen so that it includes only a single, isolated, mechanical resonance with frequency $\Omega_m$, the collective coordinate $ \hat{q}_{eff}$ becomes practically indistinguishable from the position operator $\hat{q}$ of the
selected membrane mode, and a single mechanical resonator description applies. When the term associated with the driving laser is included, the multimode Hamiltonian of Eq.~(\ref{eq:htot}) reduces just to single-mode-single-resonator Hamiltonian of Eq.~(\ref{eq:Ham-optomech}), commonly adopted for the membrane-in-the-middle setup, provided that we identify the position-dependent frequency $\omega(\hat{q})$ with its linear expansion,
\begin{equation}\label{eq:pos-dep-freq}
\omega(\hat{q})=\omega_0-g\hat{q}.
\end{equation}
In fact, the present treatment is valid only when the membrane is placed out of the nodes and antinodes of the cavity field, where the linear expansion of Eq.~(\ref{eq:pos-dep-freq}) applies. If instead the membrane is placed exactly at a node or at an antinode, the resulting radiation pressure of the cavity field $P(x,y)$ becomes equal to zero, implying $g_i=0$. In such a case the optomechanical interaction becomes dispersive and is no more associated with radiation pressure.

\section{Stability conditions}

The stationary state of the cavity-membrane system under study is stable if and only if all the eigenvalues of the drift matrix $A$ have negative real part. This condition can be re-expressed in equivalent form using the Routh-Hurwitz criteria \cite{grad}, yielding the following \emph{three} independent, nontrivial stability conditions:
\begin{widetext}
\begin{eqnarray}
 s_0 &=& \Omega_m^2\gamma_m +2 \kappa_T(q_s)\left[\Delta^2+\left[\gamma_m+ \kappa_T(q_s)\right]^2\right]+h\gamma_m \Omega_m-\Omega_m G\Gamma >0, \label{eq:stab0}\\
 s_{1} &=& 2\gamma _{m}\kappa_T(q_s) \left\{ \left[ \kappa_T(q_s)^{2}+\left(\Omega _{m}-\Delta \right)^{2}\right] \left[\kappa_T(q_s)^{2}+\left( \Omega _{m}+\Delta\right)^{2}\right]  \right. \nonumber \\
&&\left. +\gamma _{m}\left[ \left[ \gamma _{m}+2\kappa_T(q_s)\right] \left[\kappa_T(q_s)^{2}+\Delta ^{2}\right] +2\kappa_T(q_s)\Omega _{m}^{2}\right] \right\}+\Delta \Omega _{m}G^{2}\left[ \gamma _{m}+2\kappa_T(q_s)\right] ^{2} \nonumber \\
&& +\left\{\Omega_m^2\gamma_m +2 \kappa_T(q_s)\left[\Delta^2+\left[\gamma_m+ \kappa_T(q_s)\right]^2\right]\right\}\left[\Omega_m G \Gamma+2 h \Omega_m \kappa_T(q_s)\right]\nonumber \\
&&+\Omega_m\left(h \gamma_m-G \Gamma\right)\left[\Omega_m G \Gamma+2\kappa_T(q_s)\Omega_m (h+\Omega_m)+\gamma_m \left[ \kappa_T(q_s)^{2}+\Delta ^{2}\right]\right]\nonumber \\
&& -\Omega_m \left[2\kappa_T(q_s)+\gamma_m\right]^2\left[h\left[ \kappa_T(q_s)^{2}+\Delta ^{2}\right]+\kappa_T(q_s)G \Gamma \right] >0, \label{stab1}\\
s_{2} &=& \left(\Omega _{m}+h\right)\left[ \kappa_T(q_s)^{2}+\Delta ^{2}\right] -G^{2}\Delta +\kappa_T(q_s)G \Gamma  >0. \label{stab2}
\end{eqnarray}
\end{widetext}
These stability conditions are more involved with respect to the standard optomechanical case without membrane absorption and mechanical frequency shift ($h=\Gamma=0$), discussed in \cite{amo,prl07}. In particular, there is the additional stability condition given by $s_0$, which is instead always satisfied when $h=\Gamma=0$: its violation corresponds to an ``antistable'' situation, i.e., when the vibrational resonance frequency becomes imaginary.
We recall that the violation of the second condition, $s_{1}< 0$, corresponds to the emergence of a self-sustained oscillation regime where the effective mechanical damping rate vanishes. In this regime, the laser energy
leaks into the sidebands and also feeds the coherent oscillations of the membrane vibrational mode. A complex
multistable regime can emerge as described in \cite{harris06}. The violation of the second condition $s_{2}< 0$ instead corresponds to the emergence of a bistable behavior observed in \cite{dorsel}.

\end{appendix}

                                                                                            %

\end{document}